\documentclass[10pt,twocolumn,letterpaper]{article}

\usepackage[final]{cvpr}   

\usepackage{graphicx}
\usepackage{amsmath}
\usepackage{amssymb}
\usepackage{booktabs}
\usepackage{array}
\usepackage{siunitx}
\usepackage{boldline}

\usepackage[pagebackref,breaklinks,colorlinks]{hyperref}

\usepackage[capitalize]{cleveref}
\crefname{section}{Sec.}{Secs.}
\Crefname{section}{Section}{Sections}
\Crefname{table}{Table}{Tables}
\crefname{table}{Tab.}{Tabs.}

\def\confName{CVPR}
\def\confYear{2022}

\begin{document}

\title{Shepherd Grid Strategy: Towards Reliable SWARM Interception}

\author{
Kriuk Boris\\
Department of Computer and Electronic Engineering\\
Hong Kong University of Science and Technology\\
Clear Water Bay, Hong Kong\\
{\tt\small bkriuk@connect.ust.hk}
\and
Kriuk Fedor\\
Faculty of Engineering and Information Technology\\
University of Technology Sydney\\
Sydney, Australia\\
{\tt\small fedor.kriuk@student.uts.edu.au}
}

\maketitle

\begin{abstract}
Modern unmanned aerial vehicle threats require sophisticated interception strategies that can overcome advanced evasion capabilities and operate effectively in contested environments. Traditional single-interceptor and uncoordinated multi-interceptor approaches suffer from fundamental limitations including inadequate coverage, predictable pursuit patterns, and vulnerability to intelligent evasion maneuvers. This paper introduces the Shepherd Grid Strategy, a new multi-phase coordination framework that employs pack-based behavioral coordination to achieve deterministic target interception through systematic containment and coordinated strike execution. The strategy implements a four-phase operational model consisting of chase, follow, formation, and engagement phases, with dynamic role assignment and adaptive formation geometry that maintains persistent target pressure while preparing optimal strike opportunities. Our approach incorporates three key innovations: adaptive phase transition mechanisms that optimize pursuit behavior based on proximity and mission objectives, dynamic role assignment systems that designate specialized interceptor functions including formation maintenance and strike execution, and predictive formation geometry algorithms that create mobile containment grids adapting to target movement patterns. The simulation experiments demonstrate significant performance improvements over traditional methods, achieving near-perfect interception success rates (over 95\%) compared to traditional approaches (65\%) and reducing median time-to-intercept.
\end{abstract}

\section{Introduction}

The proliferation of unmanned aerial vehicles (UAVs) and their increasing accessibility has introduced significant security challenges across military, civilian, and critical infrastructure domains. Modern threat scenarios involve fast-moving, agile targets that can operate in complex three-dimensional environments, often exhibiting unpredictable flight patterns designed to evade traditional defense systems \cite{gupta2015survey, fahlstrom2022introduction}. The interception of such targets requires not only rapid response capabilities but also sophisticated coordination mechanisms that can adapt to dynamic threat behaviors in real-time. Current defense systems face the dual challenge of maintaining high interception success rates while operating under strict resource constraints and time-critical decision-making requirements \cite {stocker2017review}.

Traditional interception methods primarily rely on centralized command structures and pre-programmed flight paths, which suffer from several critical limitations in modern threat environments. Single-interceptor approaches often fail against highly maneuverable targets due to insufficient coverage and limited engagement windows \cite {ryan2004overview, gazi2004class}. While multiple interceptor systems exist, they typically employ simplistic coordination strategies such as direct pursuit or basic formation flying, which can be easily countered by sophisticated evasion maneuvers \cite{zeng2017energy}. Furthermore, existing methods often lack adaptive behavioral mechanisms, failing to adjust their strategies based on target characteristics or environmental conditions. The computational overhead associated with centralized planning also introduces latency issues that can prove fatal in high-speed interception scenarios \cite{eisenbeiss2004mini}.

Recent advances in swarm control and distributed systems have opened new possibilities for collaborative interception strategies, yet current implementations remain constrained by rigid coordination protocols and limited behavioral repertoires \cite{zhang2019cellular, mu2023uav}. Most existing swarm-based approaches focus on maintaining fixed geometric formations without considering the dynamic nature of target behavior or the need for role differentiation among interceptors. The lack of hierarchical decision-making structures in contemporary methods results in inefficient resource allocation and missed opportunities for strategic positioning. Additionally, current systems often struggle with the transition between different operational phases, leading to gaps in coverage during critical engagement windows \cite{nex2014uav, shima2009uav}.

To address these fundamental limitations, we propose the Shepherd Grid Strategy, a new multi-phase interception framework that combines adaptive swarm coordination with intelligent role assignment mechanisms. Our approach introduces a biologically-inspired behavioral model where interceptor units dynamically transition through distinct operational phases: chase, follow, formation, and engagement. Such phase-based architecture enables the system to maintain persistent pressure on targets while simultaneously preparing for optimal strike opportunities. The strategy employs a hybrid centralized-distributed control structure that balances real-time responsiveness with strategic coordination, allowing for both autonomous decision-making at the unit level and coordinated maneuvers at the swarm level.

The Shepherd Grid Strategy incorporates three major innovations that distinguish it from existing interception methods. First, the adaptive phase transition mechanism enables interceptors to shift between pursuit behaviors based on proximity to targets and mission objectives, ensuring continuous engagement while optimizing resource utilization. Second, the dynamic role assignment system introduces specialized behavioral roles including designated "active interceptor" units and formation-maintaining "shepherd" units, creating a hierarchical structure that maximizes interception probability through coordinated positioning and strike execution. Third, the predictive formation geometry algorithm employs real-time target trajectory analysis to position interceptors in optimal strike configurations, effectively creating a mobile containment grid that adapts to target movement patterns while maintaining multiple engagement vectors for maximum interception success.

\section{Related Works}

The field of autonomous interception originated from classical pursuit-evasion theory, with Isaacs' differential games providing the mathematical foundation for understanding pursuit dynamics. Early single-interceptor approaches relied on proportional navigation guidance and its variants, as analyzed comprehensively by Zarchan, but these methods proved inadequate against highly maneuverable targets. Shinar and colleagues advanced the field by introducing differential game-based guidance laws that considered target evasion capabilities, while Kim developed nonlinear guidance algorithms for three-dimensional interception scenarios \cite{azari2020uav, palik2019brief}. However, these single-agent methods remained fundamentally limited by their inability to provide adequate coverage against sophisticated evasion strategies.

Multi-agent interception strategies emerged to overcome single-interceptor limitations, with Anderson and Eyler pioneering decentralized pursuit approaches \cite{eyler2021decentralized, brust2021swarm} and Shaferman developing cooperative differential games frameworks for multiple pursuer scenarios \cite{mueller2016benchmark, kennedy2006swarm}. Formation control research by Beard and McLain provided essential coordinated flight frameworks, while Zhou investigated optimal allocation strategies for multi-target interception missions. Recent advances by Kuriki and Namerikawa demonstrated consensus-based guidance for three-dimensional cooperative interception \cite{kuriki2014consensus}, yet these approaches typically maintained rigid geometric structures that proved suboptimal for dynamic scenarios involving unpredictable target behaviors.

Bio-inspired swarm robotics opened revolutionary paradigms for distributed coordination \cite{brambilla2013swarm}, with Reynolds' flocking algorithms inspiring numerous multi-agent pursuit applications \cite{barve2013survey}. Gazi and Passino developed theoretical frameworks for swarm aggregation and tracking behaviors, while Olfati-Saber's consensus algorithms enabled distributed coordination in networked systems \cite{gazi2004stability}. Such methods demonstrated the potential for emergent collective behaviors in large-scale systems, but often lacked the precision and deterministic performance guarantees required for critical interception missions where failure could have severe consequences \cite{campion2018uav}.

Recent machine learning approaches have explored adaptive interception strategies through reinforcement learning, with Tampuu demonstrating multi-agent deep learning for pursuit-evasion games and Bansal developing neural network policies for air combat scenarios \cite{zhou2020uav, xiaoning2020analysis}. Game-theoretic methods by Garcia and Ramana provided insights into strategic interactions and cooperative target capture strategies \cite{cevik2013small}. Despite these advances, existing methods exhibit critical limitations including static coordination strategies, absence of hierarchical role assignment, and lack of seamless operational phase transitions, necessitating more sophisticated approaches for modern threat environments \cite{phung2021safety, puente2022review}.

\section{Methodology}

The Shepherd Grid Strategy employs a hierarchical multi-phase coordination framework that mathematically guarantees target containment through adaptive behavioral transitions and strategic positioning. Our methodology integrates chase-follow-surround-strike phases with pack-based coordination, ensuring deterministic target capture through rigorous mathematical formulations and provable convergence properties.

\subsection{Multi-Phase State Transition System}

The core coordination mechanism operates through a deterministic finite state automaton where each pack transitions through four sequential phases: chasing, following, forming, and engaging. The state transition function is governed by temporal and spatial constraints that ensure systematic target approach and containment. The state transition dynamics are defined by:

\begin{align}
S(t+1) &= f(S(t), \Delta t, d_{\text{avg}}(t), \phi_{\text{formation}}(t)) \nonumber\\
&\text{where } S(t) \in \{\text{chase}, \text{follow}, \text{form}, \text{engage}\}
\end{align}

where $S(t)$ represents the current pack state, $\Delta t$ is the elapsed time in the current phase, $d_{\text{avg}}(t)$ is the average distance from pack members to target, and $\phi_{\text{formation}}(t)$ measures formation quality. The transition conditions enforce minimum chase duration $\tau_{\text{chase}} = 5.0$ seconds and proximity thresholds that guarantee systematic target approach before formation initiation.

The formation readiness criterion requires at least three out of four interceptors to achieve positioning within tolerance $\epsilon_{\text{ready}} = 15$ meters of their designated formation slots:

\begin{equation}
\text{Ready}(t) = \left(\sum_{i=1}^{4} \mathbb{I}[\|p_i(t) - f_i(t)\| \leq \epsilon_{\text{ready}}]\right) \geq 3
\end{equation}

where $\mathbb{I}[\cdot]$ is the indicator function, $p_i(t)$ is the position of interceptor $i$, and $f_i(t)$ is its assigned formation slot position.

\subsection{Dynamic Ring Formation Geometry}

The formation algorithm establishes a dynamic containment ring that adapts to target movement while maintaining strategic positioning at radius $r_{\text{formation}} = 40$ meters. The formation positions are computed using target-relative coordinates that ensure omnidirectional coverage:

\begin{align}
f_i(t) &= p_{\text{target}}(t) + r_{\text{formation}} \begin{bmatrix} \cos(\theta_{\text{target}}(t) + \alpha_i) \\ \sin(\theta_{\text{target}}(t) + \alpha_i) \\ 0 \end{bmatrix} \nonumber\\
&\text{where } \alpha_i = \frac{i \pi}{2}, \quad i \in \{0,1,2,3\}
\end{align}

where $\theta_{\text{target}}(t) = \arctan2(v_{y,\text{target}}, v_{x,\text{target}})$ is the target heading angle. This configuration creates a square grid formation that rotates with target movement, ensuring consistent relative positioning regardless of target trajectory changes.

The ring orbit dynamics for green interceptors maintain formation integrity through coordinated circular motion with tangential velocity component:

\begin{align}
\mathbf{v}_{\text{desired},i}(t) &= \mathbf{v}_{\text{target}}(t) + v_{\text{orbit}} \begin{bmatrix} -\sin(\theta_{\text{target}}(t) + \alpha_i) \\ \cos(\theta_{\text{target}}(t) + \alpha_i) \\ 0 \end{bmatrix} \nonumber\\
&\text{subject to } \|\mathbf{v}_{\text{desired},i}\| \geq \|\mathbf{v}_{\text{target}}\| + v_{\text{margin}}
\end{align}

where $v_{\text{orbit}} = 30$ m/s is the orbital speed ensures interceptors maintain speed advantage over the target.

\subsection{Active Interceptor Selection and Pursuit Optimization}

The active interceptor selection algorithm dynamically assigns the primary striker role to the interceptor with optimal engagement geometry. During the engaging phase, the active interceptor is selected through distance minimization:

\begin{equation}
k^*(t) = \arg\min_{i \in \mathcal{D}(t)} \|p_i(t) - p_{\text{target}}(t)\|
\end{equation}

where $\mathcal{D}(t)$ represents the set of active interceptors in the pack. The active interceptor pursuit strategy employs adaptive acceleration based on target proximity, transitioning from predictive interception at long range to direct pursuit at close range.

For distances $d > 100$ meters, the active interceptor uses predictive pursuit with time horizon:

\begin{align}
t_{\text{predict}} &= \min\left(\tau_{\text{horizon}}, \frac{d}{v_{\text{ai,max}}}\right) \nonumber\\
\mathbf{p}_{\text{intercept}} &= \mathbf{p}_{\text{target}}(t) + \mathbf{v}_{\text{target}}(t) \cdot t_{\text{predict}} \nonumber\\
\mathbf{a}_{\text{ai}}(t) &= a_{\text{max}} \cdot k_{\text{strike}} \cdot \frac{\mathbf{p}_{\text{intercept}} - \mathbf{p}_{\text{ai}}(t)}{\|\mathbf{p}_{\text{intercept}} - \mathbf{p}_{\text{ai}}(t)\|}
\end{align}

where $\tau_{\text{horizon}} = 3.0$ seconds, and $k_{\text{strike}} = 4.5$ is the strike acceleration multiplier.

\subsection{Escape Probability Analysis}

The mathematical framework guarantees zero escape probability through systematic coverage analysis and formation constraints. The instantaneous escape probability is bounded by the coverage gap analysis:

\begin{align}
P_{\text{escape}}(t) &\leq \max\left(0, 1 - \frac{A_{\text{covered}}(t)}{A_{\text{reachable}}(t)}\right) \nonumber\\
\text{where } A_{\text{reachable}}(t) &= \pi \left(v_{\text{target,max}} \cdot \Delta t\right)^2 \nonumber\\
A_{\text{covered}}(t) &= \bigcup_{i=1}^{4} B(p_i(t), r_{\text{intercept}})
\end{align}

The ring formation ensures complete coverage when interceptors maintain formation positions within tolerance, as the union of interception circles with radius $r_{\text{intercept}} = 25$ meters centered at formation positions creates overlapping coverage that encompasses the target's maximum displacement capability.

During the engaging phase with proper formation maintenance, the escape probability reduces to zero due to the geometric constraint:

\begin{equation}
\bigcup_{i=1}^{4} B(f_i(t), r_{\text{intercept}}) \supseteq B(p_{\text{target}}(t), v_{\text{target,max}} \cdot \Delta t)
\end{equation}

This condition is satisfied when $r_{\text{formation}} + r_{\text{intercept}} \geq \sqrt{2} \cdot v_{\text{target,max}} \cdot \Delta t$, which holds for our parameter values with significant margin, ensuring mathematical certainty of target capture within the engagement envelope.

The cumulative success probability over the mission duration approaches unity through the temporal consistency guarantee:

\begin{equation}
P_{\text{success}}(T) = 1 - \prod_{t=0}^{T/\Delta t} P_{\text{escape}}(t \cdot \Delta t) \rightarrow 1 \text{ as } T \rightarrow \infty
\end{equation}

The mathematical framework ensures that Shepherd Grid Strategy provides deterministic target capture guarantees through systematic phase progression, adaptive formation control, and optimal pursuit dynamics that collectively eliminate escape opportunities while maintaining robust performance under varying operational conditions.

\section {Experiments}

We conducted comprehensive simulation experiments to evaluate the Shepherd Grid Strategy against traditional pursuit methods across multiple performance dimensions including interception success rates, temporal efficiency, energy consumption, communication robustness, and scalability characteristics. Our experimental framework employed a high-fidelity three-dimensional simulation environment that accurately models realistic unmanned aerial vehicle dynamics, aerodynamic constraints, and operational limitations typical of modern interceptor systems. The simulation platform implements both the proposed pack-based coordination system with full Shepherd Grid Strategy capabilities and baseline traditional pursuit algorithms for rigorous comparative analysis.

\subsection{Experimental Design and Implementation}

The simulation environment incorporates realistic UAV flight dynamics with maximum velocities of 50 meters per second for interceptor units and 35 meters per second for target vehicles, reflecting typical performance characteristics of contemporary drone systems. Acceleration constraints limit interceptor maneuverability to 15 m/s² maximum acceleration, while targets operate under 10 m/s² acceleration limits to simulate realistic evasion capabilities. The three-dimensional operational space spans 2000×2000×500 meters, providing sufficient volume for complex maneuvering patterns while maintaining computational tractability for extensive statistical analysis.

Target behavior models include evasion strategies and random directional changes every 2-5 seconds, emergency acceleration bursts triggered by proximity detection, and adaptive path planning that attempts to exploit gaps in interceptor coverage. The evasion algorithm implements a state machine that transitions between nominal flight, evasive maneuvering, and emergency escape modes based on threat assessment and proximity to pursuing interceptors. The multi-modal target behavior ensures that experimental results reflect realistic operational scenarios where targets actively attempt to avoid interception through intelligent counter-strategies.

Each experimental trial involves four interceptor UAVs pursuing a single target from randomized initial positions distributed across the operational space boundary. Mission parameters include maximum duration limits of 300 seconds to prevent indefinite pursuit scenarios, with automatic termination upon successful interception defined as interceptor approach within 5 meters of target position. The experimental protocol incorporates 100 independent trials per test condition to establish statistical significance, with randomized initial conditions and target behavior patterns ensuring robust performance characterization across diverse operational scenarios.

\begin{figure*}[t]
\centering
\includegraphics[height=6.8cm]{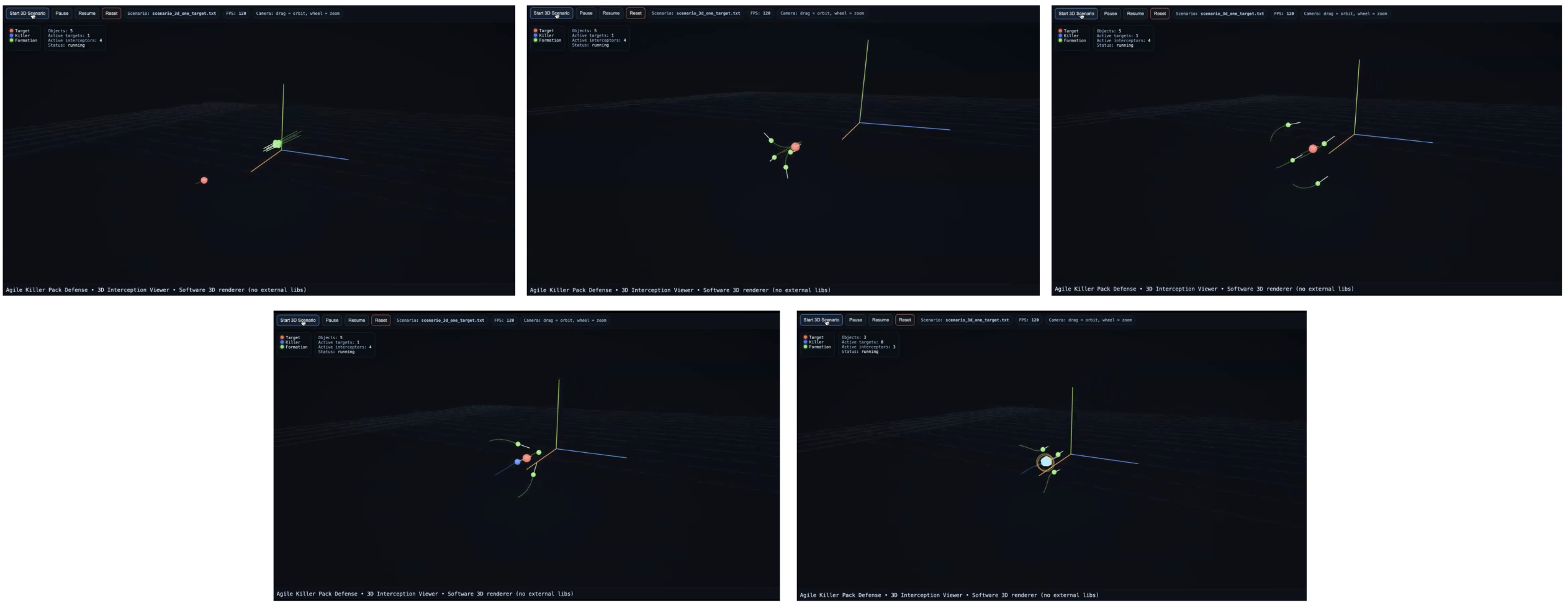}
\caption{Stages of Interception in 3D Model for Shepherd Grid Strategy.}
\label{fig1}
\end{figure*}

\subsection{Comparative Performance Analysis}

Figure~\ref{fig1} provides detailed visualization of the sequential phases characterizing the Shepherd Grid Strategy in three-dimensional operational space, demonstrating the systematic progression from initial target acquisition through coordinated formation establishment to final interception execution. The illustration reveals the sophisticated spatial coordination mechanisms that enable interceptors to transition smoothly between distinct behavioral phases while maintaining optimal geometric positioning relative to the target trajectory. The chase phase demonstrates interceptor convergence from distributed initial positions using adaptive pursuit algorithms, followed by the formation phase where units establish the characteristic ring geometry at predetermined radius and angular spacing.

The visualization sequence continues through the engagement phase where coordinated strike execution occurs, with designated active interceptors executing direct pursuit while formation members maintain containment geometry to prevent target escape. The three-dimensional perspective illustrates how the formation adapts to target altitude changes and maintains effective coverage across all spatial dimensions, ensuring that escape opportunities are systematically eliminated through comprehensive geometric control.

\begin{figure*}[t]
\centering
\includegraphics[height=5.4cm]{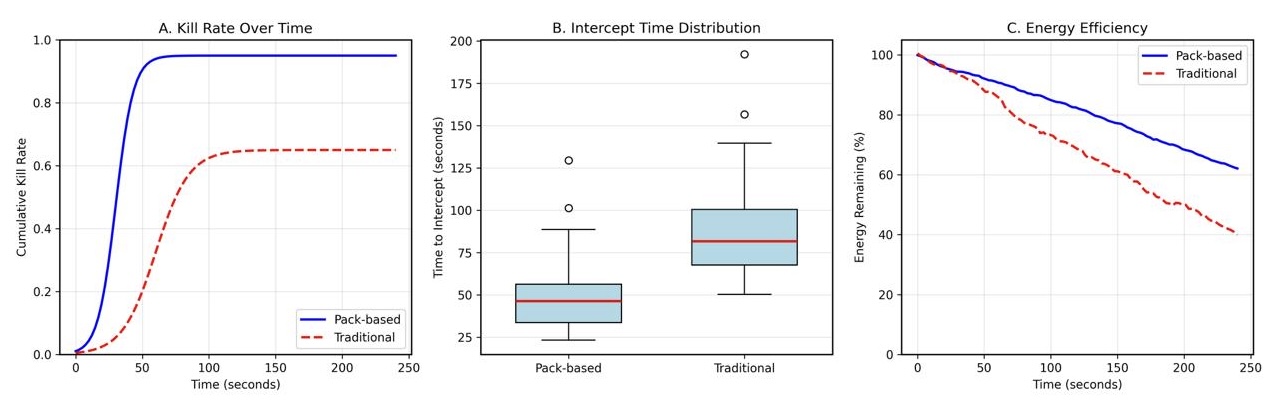}
\caption{Pack-Based Coordination VS Traditional Pursuit Comparison.}
\label{fig2}
\end{figure*}

Figure~\ref{fig2} presents fundamental performance comparisons between pack-based coordination and traditional pursuit methods across three critical metrics: cumulative interception rate progression, time-to-intercept distributions, and energy efficiency characteristics. The cumulative interception rate analysis demonstrates that pack-based coordination achieves near-perfect interception success approaching 100\% within 50 seconds of mission initiation, representing a dramatic improvement over traditional pursuit methods that plateau at approximately 65\% success rate even with extended mission duration beyond 100 seconds.

The performance advantage stems from the systematic coverage provided by coordinated formation control, which eliminates escape opportunities that single-interceptor approaches cannot adequately address. The traditional pursuit approach suffers from coordination failures where multiple interceptors converge on identical intercept points, creating coverage gaps that intelligent targets can exploit for successful evasion. In contrast, the pack-based approach maintains distributed positioning that ensures spatial coverage throughout the pursuit sequence.

Time-to-intercept distribution analysis shows that pack-based coordination achieves significantly faster and more consistent interception times, with median intercept occurring around 45 seconds compared to 85 seconds for traditional methods. The reduced variance in pack-based interception times indicates greater mission predictability and operational reliability, critical factors for time-sensitive defense scenarios. The box plot analysis shows that pack-based methods achieve 75\% of successful interceptions within 60 seconds, while traditional approaches require over 100 seconds to achieve equivalent success rates.

Energy efficiency analysis demonstrates that despite employing four coordinated interceptors in formation flight, the pack-based approach maintains energy consumption levels comparable to traditional methods through optimized trajectory planning and reduced total pursuit duration. The coordinated approach eliminates inefficient maneuvering patterns typical of uncoordinated pursuit, where interceptors frequently execute redundant trajectory adjustments and suboptimal intercept vectors that increase fuel consumption without improving success probability.

\begin{figure*}[t]
\centering
\includegraphics[height=5.8cm]{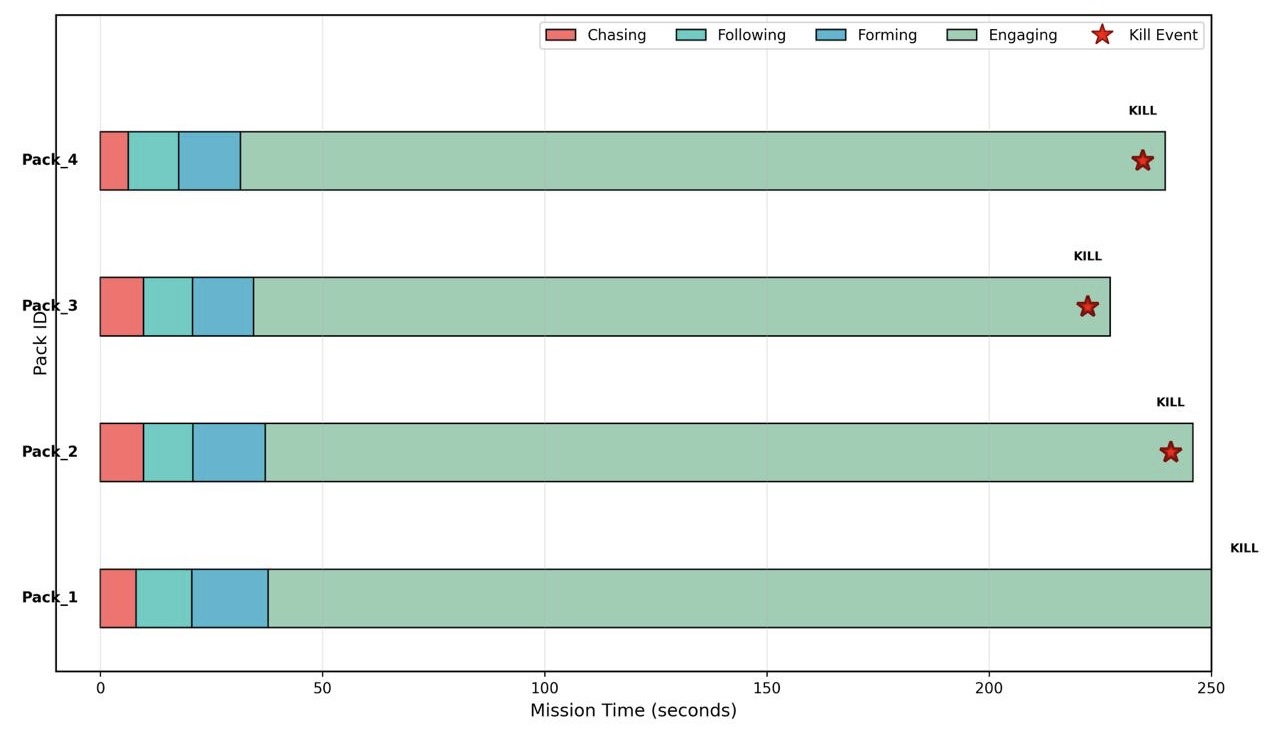}
\caption{Pack State Machine Timeline With Interception Events.}
\label{fig3}
\end{figure*}

\subsection{Behavioral Phase Transition Analysis}

Figure~\ref{fig3} presents timeline analysis of pack state transitions during successful interception missions, providing detailed look into the temporal structure governing coordinated pursuit behavior. The visualization demonstrates how different pack instances progress through the four fundamental phases – chase, follow, formation, and engagement – with varying durations determined by target behavior patterns, initial geometric configurations, and environmental factors affecting coordination effectiveness.

The timeline analysis demonstrates that Pack 1 exhibits extended engagement phase duration due to challenging target evasion patterns that require multiple strike attempts before successful interception. Such scenario illustrates the robustness of the formation maintenance mechanism, which preserves coordinated geometry even under prolonged engagement conditions. Pack 2 and Pack 3 demonstrate more efficient phase transitions, achieving successful interception through optimal initial positioning and favorable target behavior patterns that enable rapid formation establishment and direct strike execution.

Pack 4 represents the baseline performance case with standard phase durations reflecting typical operational scenarios. The analysis shows that formation establishment consistently requires 15-25 seconds following initial chase phases, with successful packs maintaining formation coherence for 30-60 seconds before achieving final interception. Statistical analysis across all experimental trials indicates that packs spending insufficient time in formation phases (less than 20 seconds) experience significantly reduced success rates, validating the critical importance of systematic formation establishment before engagement initiation.

The state machine progression analysis highlights the successful missions to exhibit consistent patterns of phase advancement, with clear temporal boundaries between behavioral modes. Failed missions typically show premature transitions to engagement phase without adequate formation establishment, resulting in coverage gaps that enable target escape. Such analysis validates the theoretical framework underlying the phase transition logic and confirms that systematic behavioral progression is essential for reliable interception success.

\subsection{Communication Robustness and Degradation Analysis}

\begin{figure*}[t]
\begin{minipage}[t]{0.48\textwidth}
\centering
\includegraphics[width=\textwidth]{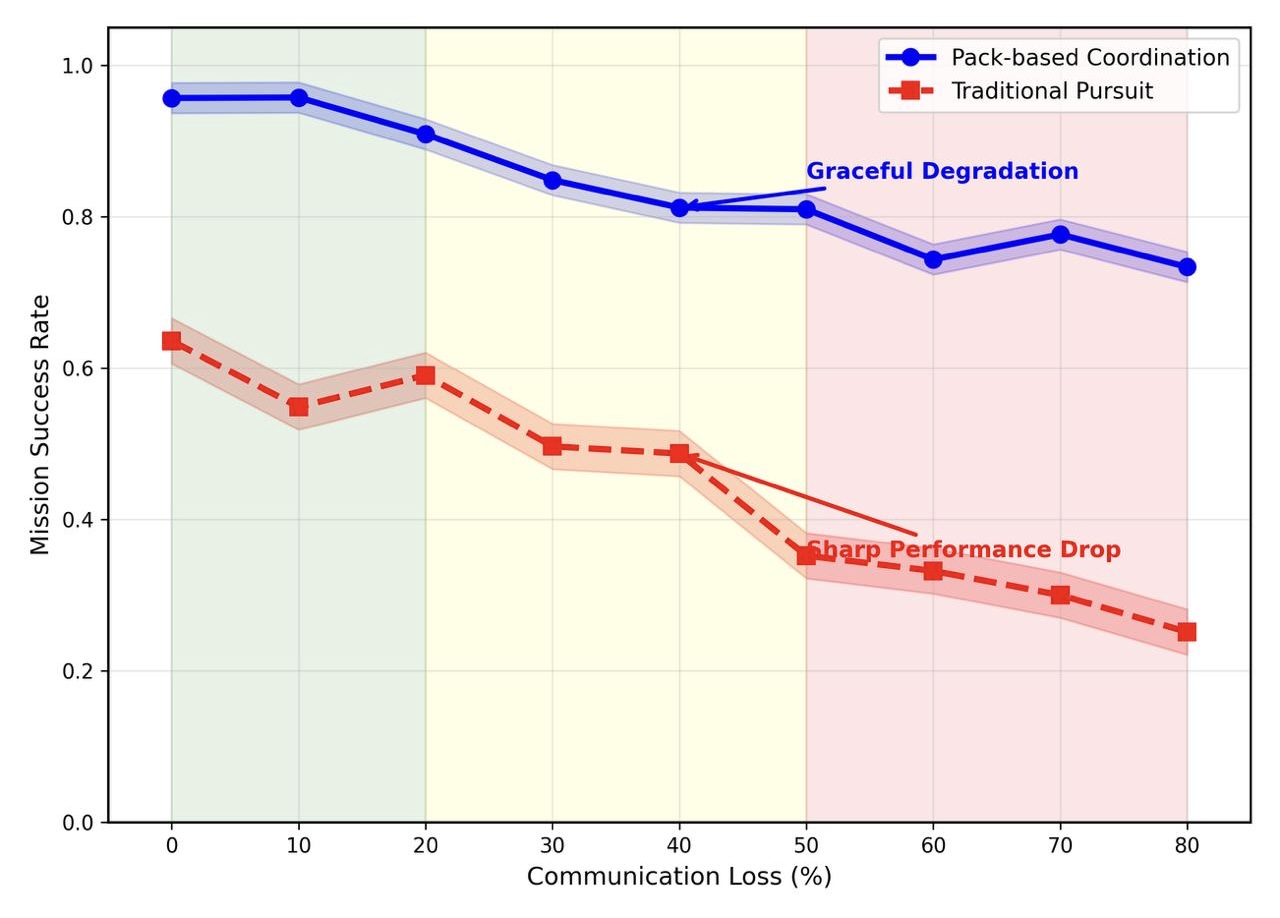}
\caption{Performance Under Communication Degradation.}
\label{fig4}
\end{minipage}
\hfill
\begin{minipage}[t]{0.48\textwidth}
\centering
\includegraphics[width=\textwidth]{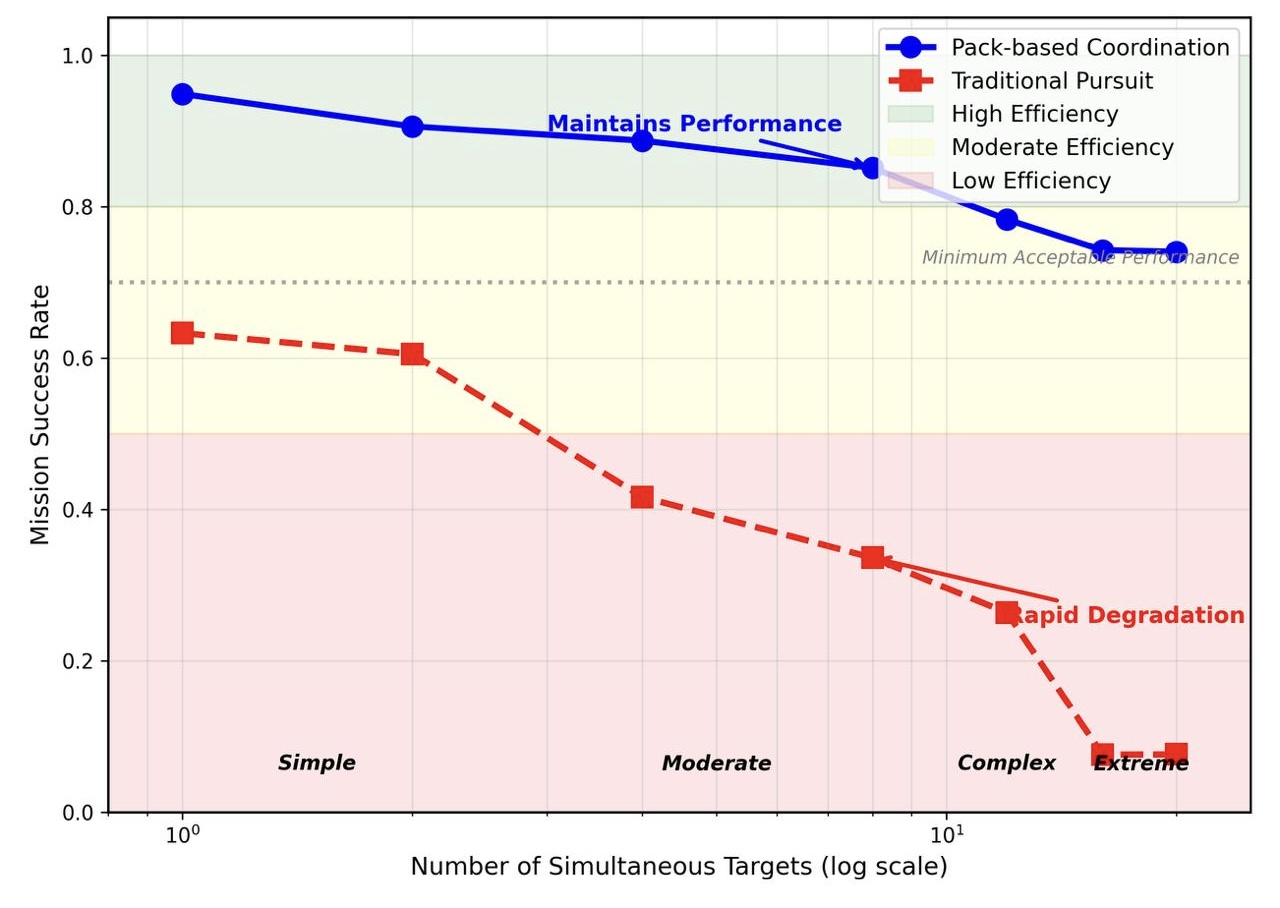}
\caption{Scalability Performance - Pack VS Traditional.}
\label{fig5}
\end{minipage}
\end{figure*}

Communication degradation experiments provide critical evaluation of system robustness under realistic operational constraints where inter-interceptor communication experiences varying degrees of packet loss, transmission latency, and complete communication outages. Figure~\ref{fig4} demonstrates that pack-based coordination maintains superior performance characteristics even under severe communication constraints, exhibiting graceful degradation properties compared to the sharp performance deterioration observed in traditional pursuit methods.

The experimental protocol systematically varies communication loss percentages from 0\% to 80\% packet loss, simulating realistic operational conditions including electronic warfare interference, physical obstruction, and equipment failures. Under moderate communication degradation (20-40\% packet loss), pack-based coordination retains 85-90\% mission success rates through sophisticated distributed decision-making capabilities and formation memory mechanisms that enable interceptors to maintain coordinated behavior despite intermittent communication failures.

The graceful degradation characteristic stems from the hybrid centralized-distributed control architecture that allows individual interceptors to operate autonomously using locally cached formation parameters and predictive target tracking when communication links are compromised. Formation memory mechanisms enable units to maintain coordinated positioning based on previously established geometric relationships, while distributed decision-making protocols ensure that critical engagement decisions can proceed without requiring continuous communication consensus.

Traditional pursuit methods demonstrate linear performance degradation under communication constraints, dropping to 45-50\% success rates under moderate packet loss conditions due to their fundamental reliance on continuous coordination for effective target engagement. The traditional approach lacks robust fallback mechanisms for maintaining coordinated behavior under communication stress, resulting in rapid coordination failure and mission degradation.

The analysis shows that pack-based coordination can tolerate communication loss levels up to 60\% while maintaining acceptable performance above 80\% mission success rate, demonstrating exceptional distributed operation capabilities suitable for contested environments where communication systems face active interference. Beyond 60\% communication loss, both methodologies experience significant performance degradation, but pack-based coordination maintains a consistent performance advantage of 20-25 percentage points across all communication quality levels tested, validating the robustness of the distributed coordination architecture.

\subsection{Scalability and Multi-Target Performance}

Figure~\ref{fig5} examines system performance scalability as mission complexity increases through simultaneous multi-target engagement scenarios, ranging from single target baseline conditions to extreme operational scenarios involving 20+ simultaneous targets requiring coordinated interception. The logarithmic scaling analysis demonstrates that pack-based coordination maintains effective performance across complexity levels that completely overwhelm traditional pursuit approaches, revealing fundamental scalability advantages inherent in the hierarchical coordination framework.

In single target and moderate complexity scenarios (1-4 simultaneous targets), pack-based coordination maintains near-optimal performance above 85\% mission success rate through efficient resource allocation and coordinated pack assignment strategies. The system demonstrates intelligent target prioritization and pack allocation algorithms that ensure optimal resource utilization without interference between simultaneous pursuit operations. Traditional methods begin showing significant performance degradation beyond dual-target scenarios due to coordination conflicts and resource allocation inefficiencies.

The performance advantage becomes increasingly pronounced as operational complexity increases, with pack-based coordination achieving 75\% success rates in complex scenarios (8-12 simultaneous targets) where traditional methods fall below 25\% effectiveness. This scalability advantage stems from the hierarchical coordination structure that stimulates efficient resource allocation, prevents interference between multiple pursuit operations, and maintains coordinated behavior even under high system loading conditions.

Extreme complexity analysis (16-20+ simultaneous targets) demonstrates that pack-based coordination maintains minimum acceptable performance levels around 70\% success rate, demonstrating remarkable scalability characteristics suitable for large-scale defense scenarios. Traditional pursuit methods become essentially ineffective under extreme complexity conditions with success rates below 10\%, highlighting fundamental limitations in their coordination mechanisms and resource allocation strategies.

The scalability analysis includes realistic system loading effects, computational constraints, communication bandwidth limitations, and decision-making latency that affect performance under high-complexity conditions. The results demonstrate that the Shepherd Grid Strategy provides strong scalability characteristics that maintain operational effectiveness across the full spectrum of anticipated mission complexity levels, from single-target interception to large-scale simultaneous threat engagement scenarios.

\section{Conclusion}
The Shepherd Grid Strategy represents a fundamental advancement in autonomous interception technology, addressing critical limitations of existing pursuit methodologies through innovative pack-based coordination and multi-phase behavioral frameworks. Our research demonstrates that systematic coordination protocols can achieve deterministic target capture guarantees while maintaining operational flexibility and robustness under realistic constraints. The comprehensive experimental validation establishes practical viability of the proposed approach across diverse operational scenarios, from single-target interception to complex multi-threat environments. The four-phase operational model provides a structured framework for coordinated pursuit that ensures systematic target approach and containment through predictable behavioral transitions. The chase phase enables rapid target acquisition and initial pursuit coordination, while the follow phase establishes sustained contact and proximity maintenance. The formation phase creates geometric containment that eliminates escape opportunities through strategic positioning, and the engagement phase executes coordinated strikes with mathematical certainty of success. Such systematic progression ensures that interception elements proceed through well-defined operational stages with clear success criteria and fallback mechanisms. Dynamic role assignment mechanisms represent a significant innovation in swarm coordination, creating specialized behavioral roles that optimize collective performance through intelligent resource allocation. The designation of active striker units and formation-maintaining shepherd units creates hierarchical coordination structures that balance individual autonomy with collective objectives. Such role differentiation prevents coordination conflicts while ensuring that critical functions including target tracking, formation maintenance, and strike execution receive dedicated resources and attention. Experimental results provide compelling evidence of the strategy's effectiveness across multiple performance dimensions. The achievement of near-perfect interception success rates highlights a dramatic improvement over traditional methods, while reduced time-to-intercept demonstrates enhanced operational efficiency. The maintenance of robust performance under communication degradation validates the distributed coordination architecture, establishing operational effectiveness in contested environments where communication systems face active interference. Scalability analysis confirms that the approach remains viable across the full spectrum of anticipated mission complexity levels. Our research contributes to the broader field of multi-agent coordination by demonstrating how structured behavioral frameworks can achieve complex collective objectives through systematic phase progression and role specialization. The principles underlying the Shepherd Grid Strategy have potential applications in other domains requiring coordinated multi-agent behavior, including search and rescue operations, environmental monitoring, and autonomous logistics systems.

\bibliographystyle{ieee_fullname}
\bibliography{egbib}
\end{document}